\documentclass[12pt]{article}
\usepackage{graphicx}
\begin{document}

\begin{center}
{\Large \bf Lens optics and the continuity problems of the $ABCD$ matrix}

\vspace{14mm}

S. Ba{\c s}kal \footnote{Electronic address:
baskal@newton.physics.metu.edu.tr}  \footnote {Corresponding author}\\
Department of Physics, University of Maryland,\\
College Park, Maryland 20742, USA\\
Department of Physics,
Middle East Technical University,\\
06531 Ankara, Turkey \footnote{Permanent address}
\vspace{7mm}

Y. S. Kim \footnote{Electronic address: yskim@umd.edu}\\
Department of Physics, University of Maryland,\\
College Park, Maryland 20742, USA

\end{center}

\vspace{6mm}

\begin{abstract}
Paraxial lens optics is discussed to study the continuity properties
of the $ABCD$ beam transfer matrix.  The two-by-two matrix for the
one-lens camera-like system can be converted to an equi-diagonal form by a scale
transformation, leaving the off-diagonal elements
invariant.  It is shown that the matrix remains continuous during the
focusing process, but this transition is not analytic.   However, its
first derivative is still continuous, which leads to the concept of
``tangential continuity.''  It is then shown that this tangential continuity
is applicable to $ABCD$ matrices pertinent to periodic optical systems, where
the equi-diagonalization is achieved by a similarity transformation
using rotations.  It is also noted that both the scale transformations and
the rotations can be unified within the framework of Hermitian
transformations.\\[3mm]


\end{abstract}
\newpage

\section{Introduction}\label{intro}

The two-by-two $ABCD$ matrix  is a very useful tool both
in ray and Gaussian beam optics and has many interesting mathematical
aspects rendering a good understanding of the physical system that
it describes.   The converse of the procedure is also true, in the
sense that the properties of the optical system guides us to explore
hidden mathematical features of the matrix.  In this article, our
aim is to investigate in both directions.

The $ABCD$ or in particular the ray transfer matrices are diagonalized
for many purposes.~\cite{marhic95,bastiaans06,bastiaans07}
However, diagonalization may not always be possible.  It has been shown
in our earlier article thatit can always be transformed to a matrix with
equi-diagonal elements by making a similarity transformation by means of
rotations.~\cite{bk13mop,bk10jmo}

For periodic systems, the calculation of the resultant matrix, that
would otherwise be obtained by multiplying a cascaded sequence
of matrices, reduces to taking the n$^{th}$ power of the $ABCD$
matrix.  Equi-diagonalization by similarity transformations relieves the
the burden of taking powers by multiplying the number of the cycles with
the matrix.  This procedure had been exemplified by a laser cavity
resonator, consisting of two identical concave mirrors.   Another
instance of a periodic system is multilayer optics, where the $ABCD$
matrix governs the two component transverse electric field  propagating
successively from one medium to another.

Also in our earlier paper~\cite{bk10jmo}, it has been explained that the
equi-diagonal form can be written as a similarity transformation of a
rotation, of a triangular, or of a squeeze matrix and  the equi-diagonalized
$ABCD$ matrix can be expressed in an exponential form with one angular
parameter and two linearly independent generators.   The resulting matrix
has been classified in connection with the values of this angular parameter.   It has been
shown that each matrix belonging to distinct classes can be unified in a
single expression having several branches.

Here, we shall elaborate on the issue of focusing of the image in a one
lens camera-like optical system, whose mathematical formulation
leads to the vanishing of the upper right component of the $ABCD$ matrix.
We shall primarily be investigating the continuity of this focusing
process.  Since, equi-diagonalization by rotations does not leave
the upper right component of the $ABCD$ matrix invariant, we shall
introduce a group of transformations to achieve the same purpose while
keeping the focal condition intact.  We will show that focusing of the
image in such a camera-like system is a "tangentially continuous" process.
Having guided by this example, the continuity of the transitions between
the branches of the $ABCD$ matrix will also be demonstrated.

It is also pointed out that such a continuity property occurs in the
$ABCD$ matrices applicable to periodic systems such as laser resonators
and multilayer optics, where the $ABCD$ matrix is brought to an
equi-diagonal form by a similarity transformation.

In Sec. \ref{lenso}, the one-lens system is studied in detail.  The
two-by-two matrix is first brought to an equi-diagonal form.   Since
we are interested in keeping the off-diagonal elements invariant, the
transformation is achieved by a scale transformation on the diagonal
elements.  It is shown that the matrix remains continuous during the
focusing process, but the continuation is not analytic.  Yet, its
first derivative is continuous. This leads to the concept of
tangential continuity.

In Sec. \ref{conti}, we study the $ABCD$ matrix which can be converted to
an equi-diagonal form by a similarity transformation using rotations.
While the one-lens system matrix is equi-diagonalized by a scale transformation,
the tangential continuity is still a prevailing
concept in discussing the nature of the continuity for this type of an
$ABCD$ matrix.

\par

\section{Lens Optics}\label{lenso}

Although problems involving the matrix formulation of paraxial ray optics
containing an object,  one lens and an image seem to be simple,
they provide valuable insight for the properties of more
general and diverse cases. The purpose of this section to analyze the
focusing of the image in a simple one lens camera-like system, and discuss
the continuity of this process, which in turn will lead us
to investigate the properties of the most general two-by-two ray matrix.

\subsection{Equi-diagonalization and the focal condition}
A simple optical arrangement for a paraxial ray consists of a lens
with focal length $f$ and the propagation of the ray by an
amount $d$.~\cite{saleh07}
The lens matrix is given by
\begin{equation}
\pmatrix{1 & 0 \cr -1/f & 1} ,
\end{equation}
and a translation of the ray is expressed by the matrix
\begin{equation}\label{tri01}
\pmatrix{1 & d \cr 0 & 1} .
\end{equation}

If the object and the image are $d_{1}$ and $d_{2}$ distances away
from the lens respectively, the system is described by
\begin{equation}\label{lens01}
\pmatrix{1  & d_{2} \cr 0 & 1} \pmatrix{1  & 0 \cr -1/f  & 1 }
\pmatrix{1  & d_{1} \cr 0 & 1} .
\end{equation}
The multiplication of these matrices leads to
\begin{equation}\label{lens02}
\pmatrix{1 - d_{2}/f  & d_{1} + d_{2} - d_{1} d_{2}/f
      \cr - 1/f & 1 - d_{1}/f} .
\end{equation}
The image becomes focused
when the upper right element of this matrix vanishes,~\cite{bk03} i.e.,
\begin{equation}\label{focus01}
{ 1 \over d_{1}} + { 1 \over d_{2}} = { 1 \over f}.
\end{equation}
With the inclusion of the focal condition such an optical arrangement
will be called as a "one-lens camera-like" system.

It is shown that the matrix remains continuous during the
focusing process, but the transition is not analytic.

\par

The matrix of Eq.(\ref{lens02}) can be equi-diagonalized
by making a similarity transformation:
\begin{equation}\label{lens03}
\pmatrix{1 &  \delta  \cr 0 & 1}
\pmatrix{1 - d_{2}/f  &  d_{1} + d_{2} - d_{1}d_{2}/f
     \cr -1/f & 1 - d_{1}/f}
\pmatrix{1 &  -\delta \cr 0 & 1},
\end{equation}
where
$$
\delta = \frac{d_{1} - d_{2}}{2} ,
$$
which  becomes
\begin{equation}\label{lens04}
\pmatrix{1-\frac{1}{2f}\left(d_{1} + d_{2}\right)  &
d_{1} + d_{2} -\frac{1}{4f}\left(d_{1} + d_{1}\right)^2
\cr -\frac{1}{f} & 1-\frac{1}{2f}\left(d_{1} + d_{2}\right)} .
\end{equation}
This matrix is equi-diagonal, and it is achieved
by making a similarity transformation using translations.
Translations similar to this process has been used earlier when
dealing with multilayer problems.~\cite{bk13mop,gk03}
In lens optics, the focal condition is the upper-right
element to be zero.  We observe that equi-diagonalization obtained as
above does not leave the upper-right element invariant.
It is also already known to us that this is also not possible through
rotations as we had presented in ~\cite{bk10jmo}.  Therefore, we are
interested in an equi-diagonalization method that will leave the
off-diagonal element invariant.

\subsection{Focusing as the Tangential Continuity}\label{focus}
Let us go back to the matrix of Eq.(\ref{lens02}).  First,
consider the case where both $d_{1}$ and $d_{2}$ are larger than $f$,
which is the case for camera optics.  It is more convenient to deal
with the negative of this matrix with positive diagonal elements.
Let us use the variables
\begin{equation}
  x_{1} =  d_{1}/f ,  \qquad  x_{2} =  d_{2}/f
\end{equation}
where, we are measuring the distance in units of the focal length.
Then, this camera matrix becomes
\begin{equation}
\pmatrix{x_{2} - 1  &  \chi \cr
 1 & x_{1} - 1}
\end{equation}
with
\begin{equation}\label{chiy}
\chi= x_{1}x_{2} - x_{1} - x_{2} .
\end{equation}

By making a scale transformation
\begin{equation}\label{htr1}
\pmatrix{b & 0 \cr 0 & 1/b}
\pmatrix{x_{2} - 1 &  \chi  \cr 1 & x_{1} - 1}
\pmatrix{b & 0 \cr 0 & 1/b} ,
\end{equation}
with $$
b = \left( {x_{1} - 1 \over x_{2} - 1 } \right)^{1/4} ,
$$
the camera matrix receives equal diagonal elements
\begin{equation}\label{lens11}
\pmatrix{\sqrt{1 + \chi}  &   \chi \cr 1 &  \sqrt{1 + \chi}} ,
\end{equation}
where
\begin{equation}
1 + \chi = (x_{1} - 1)(x_{2} - 1).
\end{equation}

When $\chi$ is negative, the camera matrix of Eq.(\ref{lens11}) can be
written as
\begin{equation}\label{lens12}\label{sqxm}
\pmatrix{\sqrt{1-\xi^2}&   - \xi^2
\cr  1 & \sqrt{1-\xi^2}} ,
\end{equation}
where
$- \xi^2 =  -|\chi|$.

When $\chi$ is positive, the camera matrix should take the form
\begin{equation}\label{sqxp}
\pmatrix{\sqrt{1+\xi^2}&    \xi^2
\cr  1 & \sqrt{1+\xi^2}} ,
\end{equation}
where $\xi^{2} =  \chi $.

\begin{figure}
\centerline{\includegraphics[scale=0.60]{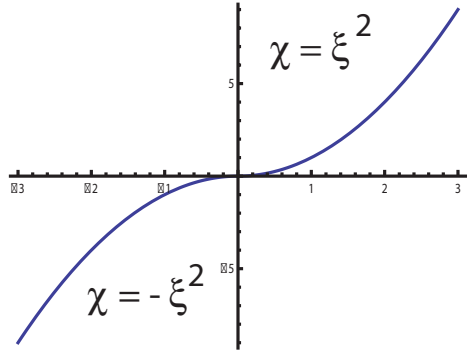}}
\caption{Tangential continuity during the focusing process.
The function takes a different form when the variable $\chi$
passes from negative to positive values. Yet the transition
is continues with continuous first derivatives.  Their
second derivatives take different forms.  Thus, it is a
tangential but not analytic continuation.}\label{xsquare}
\end{figure}

When $\chi = 0$,
\begin{equation} \label{lens16}
\frac{1}{x_{1}} + \frac{1}{x_{2}} = 1 .
\end{equation}
the camera is focused, and the matrix becomes
\begin{equation} \label{lens17}
\pmatrix{1 & 0 \cr 1 & 1} .
\end{equation}

During the focusing process, the upper right component of the camera
matrix goes from a negative value to a positive value, which is a continuous process,
as demonstrated in Fig.\ref{xsquare}.  The second derivatives is not continuous, therefore
this kind of continuity is not analytic.  Nevertheless, since the first derivatives are continuous,
the process is tangentially continuous.

If $ x_{2}$ is $x_{0}$, where
\begin{equation}\label{lens18}
x_{0} = \frac{x_{1}}{x_{1} - 1} ,
\end{equation}
the camera is focused.  If $x_{1}$ is very large, $x_{0}$ approaches
the focal length.  If $x_{1}$ is smaller than $x_{0}$, $\chi$ is
negative, and the camera matrix take the form of Eq.(\ref{lens12}).
If $x_{1}$ is larger than $x_{0}$, the camera matrix takes the
form of Eq.(\ref{sqxp}).

\subsection{Mathematical Summary}\label{math}
In this section, we were first interested in achieving the
equi-diagonalization of the two-by-two matrix without changing the
off-diagonal elements.  For this purpose, we used a Hermitian
transformation which produces a scale change on the diagonal
elements.

It was then noted that the focusing process corresponds the
transiton from Eq.(\ref{sqxm}) to Eq.(\ref{sqxp}).   Their
traces are smaller and greater than 2, respectively.  As shown
in Fig.~\ref{xsq}, this results in the ``soldering'' of two different
functions.  The the functions and their first derivatives are continuous at
the point where they are glued together and consequently the property is
termed as tangentially continuous.

In our earlier papers, we studied the $ABCD$ matrix applicable to periodic
systems such as laser cavities, where the equi-diagonalization
is achieved through a  similarity transformation by using rotations.
Since the inverse of the rotation matrix is also its Hermitian
conjugate, we can put both the scale transformation and the
rotation into one set of Hermitian trasformations.

\section{The Equi-diagonalization and the Continuity of the $ABCD$
matrix}\label{conti}

The $ABCD$ matrices are essential in understanding the propagation of light,
both in ray optics and Gaussian beam optics. Its determinant is
one for lossless systems. For paraxial ray optics, this two-by-two matrix
has real components, and in view of the condition on its determinant the
number of independent components reduces to three.

Therefore, from a group theoretical point of view, it can be represented in
terms of the symplectic group $Sp(2)$, consisting of one rotation and two
squeeze matrices, whose properties have been extensively discussed in
connection with its applicability to optics.~\cite{bk13mop}

It was noted in our earlier publications that this matrix cannot always be
diagonalized, but it can be brought to a form with equal diagonal elements.
This equi-diagonal form can serve many useful computational purposes.

We noted further that the $ABCD$ matrix can be brought to an
equi-diagonal form by a rotation.  As was noted in Sbsec.~\ref{math},
this rotation and  the scale change in lens optics  can be grouped into
a set of Hermitian transformations.

Furthermore, it was shown in one of our earlier papers that the equi-diagonalized
matrix can be expressed in an exponential form with two matrix generators
and one angular parameter.  Depending on the values of this parameter, the
matrix has four branches.~\cite{bk10jmo}  However, we noted that the
$ABCD$ matrix maintains its continuity while crossing from one branch to
another.  We noted there that it is not an analytic continutation, but
we could not go further than that. In this section, we shall conclude that
this continuity is the tangential continuity as discussed in Sec.~\ref{lenso}.

 We shall discuss the problems
of equi-diagonalization in full detail in Sec.~\ref{equi}.

\subsection{Equi-diagonalization of the $ABCD$ matrix} \label{equi}
The $ABCD$ matrix can not always be brought into a diagonal form,
but it is possible to bring it into an equi-diagonal form by rotations~\cite{bk13mop}
\begin{equation}
[abcd]_{R}=R(-\sigma/2)[ABCD]R(\sigma/2),
\end{equation}
where the $ABCD$ and the $[abcd]$ matrices are of the form
\begin{equation}\label{oabcd}
[ABCD] = \pmatrix{A & B \cr C & D}, \qquad [abcd]=\pmatrix{a & b \cr c & d},
\end{equation}
respectively, and $R(\sigma)$ is
\begin{equation}\label{rot}
\pmatrix{\cos(\sigma/2) & -\sin(\sigma/2) \cr
   \sin(\sigma/2) & \cos(\sigma/2)}
\end{equation}
is the rotation matrix.
Here $(abcd)_{R}$ denotes the equi-diagonal matrix achieved by rotations,
and $a,b,c$ and $d$ can be expressed in terms of the elements of Eq.(\ref{oabcd})
where
\begin{equation}
\tan\sigma=\frac{A-D}{B+C}.
\end{equation}

\par
However, transformations by rotations are not the only way to bring it to an equi-diagonal form.  They can also be equi-diagonalized by squeeze matrices as
\begin{equation}\label {trans01}
[abcd]_{S}=S(\gamma/2)[ABCD]S(\gamma/2),
\end{equation}
where $S(\gamma)$ is
\begin{equation}
\pmatrix{e^{\gamma/2} & 0 \cr  0 & e^{-\gamma/2}},
\end{equation}
and the squeeze parameter
\begin{equation}
e^{\gamma} = \sqrt{D/A} ,
\end{equation}
with
\begin{equation}
a=d=\sqrt{AD}, \quad  b=B,  \quad c=C ,
\end{equation}
if $A$ and $B$ have the same sign.

We now have two different transformation matrices. One is the
rotation matrix and the other is the squeeze matrix.   It is possible to
accommodate both in a single form by a Hermitian
transformation
\begin{equation}\label{hermit11}
                L[ABCD]L^{\dagger} .
\end{equation}
where $L^{\dagger}$ is the Hermitian conjugate of $L$.
The rotation matrix is antisymmetric and its Hermitian conjugate is its inverse.  Thus,
it is a similarity transformation.  The squeeze matrix is symmetric, and it is invariant
under the Hermitian conjugation.  The Hermitian transformation of Eq.({\ref{hermit11}})
is not a similarity transformation.  The usage of rotations is far well known in optical
sciences compared to the usage of squeeze transformations, which are also well
established in the context of special relativity.~\cite{bk13mop,naimark64,sym13}

\par
Thus, equi-diagonalization is possible through transformations
\begin{equation}
      M [ABCD] M^{-1}, \quad  \mbox{or} \quad M[ABCD]M^{\dagger} ,
\end{equation}
where $M$ is a two-by-two matrix from the  group $Sp(2)$.  If the matrix $M$
is antisymmetric, its Hermitian conjugate is its inverse.  If it is
symmetric, it is invariant under the conjugation, but conjugation does yield
its inverse.  In either case, there is a variety of ways bringing the $ABDC$
matrix to a diagonal form.  We can choose the method depending on our purpose.

\par
Now, let us go back to the procedure from Eq.(\ref{lens03}) to Eq.(\ref{lens04}).
This is a translation operation which will place the lens exactly halfway between
the image and object. This is achieved by the similarity transformation of a
triangular matrix in the form
\begin{equation}\label{trigmat}
\pmatrix{1 & \kappa \cr  0 & 1}.
\end{equation}
On the other hand, this triangular matrix can be obtained by multiplications
of the squeeze and rotation matrices in the Sp(2) group.  This process is
known as the Iwasawa decomposition.~\cite{iwasawa49,bk01} However, this
process of equi-diagonalization does not leave the off-diagonal elements
invariant, and thus is not useful for focusing processes.

\subsection{Tangential continuity of the $ABCD$ matrix}\label{tcabcd}

It was shown also in our previous paper~\cite{bk10jmo} that the
equi-diagonal form of the $ABCD$ matrix can be written in the exponential
form
\begin{equation}\label{eq01}
[abcd] = \exp{\left[-i r (A\cos\theta + S\sin\theta)\right]},
\end{equation}
where
\begin{equation}
A = \pmatrix{0 & -i \cr i & 0}, \qquad S = \pmatrix{0 & i \cr i & 0} .
\end{equation}
Then, we have
\begin{equation}\label{eq02}
[abcd] = \exp{\left[r \pmatrix{0 & -\cos\theta +\sin\theta \cr
\cos\theta + \sin\theta & 0}\right]}.
\end{equation}

Let us next consider the new angle variable defined as
\begin{equation}
\alpha = \theta - \frac{\pi}{4} .
\end{equation}
Then the above exponential form can be written  as
\begin{equation}\label{eq03}
[abcd] = \exp{\left[\sqrt{2}r\pmatrix{0 & \sin\alpha \cr \cos\alpha & 0}\right]}.
\end{equation}
Now, the $[abcd]$ matrix is to be investigated for various values of the angle
$\alpha$.

Case i)~~The angle $\alpha $ is smaller than $0$ but larger than
$-\pi/2$:\\
Within this range of $\alpha$ the exponent of Eq.(\ref{eq03}) can be expressed as
\begin{equation}
r\sqrt{\sin(2|\alpha|)}
\pmatrix{0 & - e^{-\eta} \cr  e^{\eta} & 0}
\end{equation}
which can also be written as a similarity transformation
\begin{equation}\label{sim01}
r \,\,\sqrt{\sin(2|\alpha|)} \pmatrix{e^{-\eta/2} & 0 \cr 0 & e^{\eta/2}}
                                \pmatrix{0 & -1 \cr 1 & 0}
 \pmatrix{e^{\eta/2} & 0 \cr 0 & e^{-\eta/2}},
\end{equation}
where
\begin{equation}
e^{-\eta} = \sqrt{\tan(|\alpha|)}.
\end{equation}
It is apparent that the $[abcd]$ matrix is a similarity transformation of an exponential
\begin{equation}
[abcd] = \pmatrix{e^{-\eta/2} & 0 \cr 0 & e^{\eta/2}}
\exp{\left[\rho_{-}\pmatrix{0 & -1 \cr 1 & 0}\right]}
 \pmatrix{e^{\eta/2} & 0\cr 0 & e^{-\eta/2}}
\end{equation}
where
\begin{equation}
\rho_{-} = r \sqrt{\sin(2|\alpha|) }.
\end{equation}
After exponentiating the matrix becomes
\begin{equation}\label{eq06}
[abcd] = \pmatrix{\cos\rho_{-} & - e^{-\eta} \sin\rho_{-} \cr
    e^{\eta} \sin\rho_{-} & \cos\rho_{-}} .
\end{equation}

Now, it is possible to express the group parameters $\eta$ and $\rho_{+}$ in terms of the physical quantities of the camera like one lens system
$d_{1},d_{2}$ and $f$ as
\begin{equation}
e^{\eta}=\frac{1}{\sqrt{|\chi|}}, \qquad -(\sin \rho_{-})^2=-|\chi|
\end{equation}
where $\chi$ is given as in Eq.(\ref{chiy}).

\begin{figure}
\centerline{\includegraphics[scale=0.8]{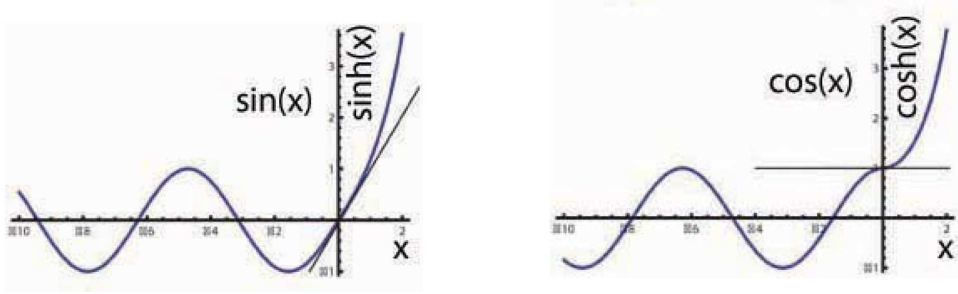}}
\caption{Transitions from $\sin(x)$ to $\sinh(x)$, and from $\cos(x)$
to $\cosh(x)$ at $x = 0$.
They are continuous transitions.  Their first derivatives are also
continuous.  However, their second and third derivatives take
different forms.  They are thus tangentially continuous
transitions. }\label{sincos}
\end{figure}

Case ii)~~The angle  $\alpha$ is larger than $0$ but less than $\pi/2$:\\
Within this range of $\alpha$ the exponent in Eq(\ref{eq03}) is expressed as
\begin{equation}\label{eq07}
r\,\,\sqrt{\sin(2\alpha)}\pmatrix{e^{-\eta/2} & 0 \cr 0 & e^{\eta/2}}
                                \pmatrix{0 & 1 \cr 1 & 0}
 \pmatrix{e^{\eta/2} & 0 \cr 0 & e^{-\eta/2}}.
\end{equation}
After exponentiating, as before, the $[abcd]$ matrix takes the form
\begin{equation}\label{eq08}
[abcd] = \pmatrix{\cosh\rho_{+} &  e^{-\eta} \sinh\rho_{+} \cr
     e^{\eta} \sinh\rho_{+} & \cosh\rho_{+}} ,
\end{equation}
where
\begin{equation}
e^{-\eta}=\sqrt{\tan(\alpha)},   \qquad \rho_{+} = r \sqrt{\sin{(2\alpha)}} .
\end{equation}

Similarly, as in the case above the group parameters are related to the
physical quantities as:
\begin{equation}
e^{\eta}=\frac{1}{\sqrt{\chi}}, \qquad (\sinh \rho_{+})^2=\chi .
\end{equation}

\par
Now, Eq.(\ref{eq06}) and Eq.(\ref{eq08}) can be combined into one exponential
form by
\begin{equation}
 \exp{\left[\rho_{\pm}\pmatrix{0 & \pm e^{-\eta} \cr e^{\eta} & 0}\right]} .
\end{equation}

\par
To examine the transition between Eq.(\ref{eq06}) and Eq.(\ref{eq08}),
Eq.(\ref{lens03}) is
expanded around small values of $\alpha$, for the cases (i) and (ii).
They become
 \begin{equation}\label{eq11}
\pmatrix{1 - |\alpha| r^2 &  -\sqrt{2} r |\alpha| \cr
          \sqrt{2} r & 1 - |\alpha| r^2} ,
\qquad
\pmatrix{1 + \alpha r^2 &  \sqrt{2} r \alpha \cr
          \sqrt{2} r  & 1 + \alpha r^2}
\end{equation}
respectively.  They attain the same form when $\alpha=0$, where both
are lower triangular matrices, with vanishing upper right components
similar to that of Eq.(\ref{lens17}), accounting for the focusing procedure.

The tangential continuity is illustrated in Fig.~\ref{sincos}, where the
transition from trigonometric to hyperbolic functions are presented,
with their common tangential lines.

\section*{Conclusion}

Our study on the equi-diagonalization and thereafter the branching property
of the $ABCD$ matrix was initiated while investigating the behavior
of light rays in periodic systems such as laser resonators and
multilayer optics, in our earlier papers.~\cite{bk10jmo}
In those papers we have also given the relations between the
group parameters and group parameters.

We have further noted that the equi-diagonal matrix can have its trace smaller
than 2, greater than 2, or equal to 2, and that the transition from one
branch to another is continuous, but we were not able to clarify the
nature of the continuity.

In this paper, we used lens optics to study this problem, and concluded
that the answer is the ``tangential continuity."  However, there are
some intricacies due to different procedures for equi-diagonalization.

For lens optics, we used the Hermtian transformation of the form
$L[ABCD]L^{\dagger}$, while the similarity transformation of the
form  $S[ABCD]S^{-1}$ is applicable to the periodic systems discussed
in Sec.~\ref{conti}.    The similarity transformation is well known
and possesses the property
\begin{equation}
\left[M[ABCD]M^{-1}\right]^n = M[ABCD]^{n}M^{-1} ,
\end{equation}
which is needed for dealing with periodic systems.

The Hermitian transformation is rare in the literature, but it
is not new.  It is applicable to Lorentz
transformations of the spacetime four-vectors in the two-by-two
representation.~\cite{bk13mop}

While these two transformations perform different mathematical operations,
transformations by rotations belong to both types.  We use rotations
as a subset of the similarity transformation in Sec.~\ref{conti}.
Since the Hermitian transformation also contains this subset, we can
include both equi-diagoanalization processes into one set of Hermitian
transformations.

The concept of tangential continuity in lens optics is directly
applicable to the $ABCD$ matrices
discussed in Sec.~\ref{conti} for periodic systems such as laser
optics and multilayer optics.  Indeed, in this paper, we have
completed our investigation of the continuity problem in the transition
from one branch of the $ABCD$ matrix to another, which was left unresolved
in our earlier paper in this journal.\cite{bk10jmo}.

\end{document}